\documentclass{PoS}
\usepackage{epsf}
\usepackage{graphicx}
\usepackage{amsmath}

\title{Exclusive  semileptonic decays of ground-state $cb$ 
baryons driven by a $c\to s,d$ quark transition}

\ShortTitle{Exclusive  semileptonic decays of ground-state $cb$ 
baryons driven by a $c\to s,d$ quark transition}

\author{\speaker{C. Albertus}\\
        Departamento de F\'\i sica Fundamental e
IUFFyM,\\ Universidad de Salamanca, E-37008 Salamanca, Spain\\
        E-mail: \email{albertus@usal.es}}

\author{E. Hern\'andez\\
        Departamento de F\'\i sica Fundamental e
IUFFyM,\\ Universidad de Salamanca, E-37008 Salamanca, Spain\\
        E-mail: \email{gajatee@usal.es}}

\author{J. Nieves\\
        Instituto de F\'\i sica Corpuscular
(IFIC), Centro Mixto CSIC-Universidad de Valencia, Institutos de
Investigaci\'on de Paterna, Aptd. 22085, E-46071 Valencia, Spain\\
        E-mail: \email{jmnieves@ific.uv.es}}

\abstract{  We evaluate  semileptonic decays of  spin-1/2
  and spin-3/2 doubly heavy $cb$ baryons. The decays are driven 
  by a $c\to s,d$ transition at
  the quark level.  We check our results for the form factors against
   heavy quark spin symmetry constraints  obtained in the
  limit of  very large heavy quark masses  and near zero recoil.}

\FullConference{Sixth International Conference on Quarks and Nuclear Physics,\\
		April 16-20, 2012\\
		Ecole Polytechnique, Palaiseau, Paris}

\newcommand{\be}{\begin{equation}}
\newcommand{\ee}{\end{equation}}
\newcommand{\bea}{\begin{eqnarray}}
\newcommand{\eea}{\end{eqnarray}}

\makeatletter
\def\slashchar#1{{\mathpalette\c@ncel{#1}}} 
\makeatother

\begin{document}

\section{Semileptonic decay widths}
\label{sect:dwff}
The total decay width for semileptonic $c\to l$ transitions, with
$l=s,d$, is given by 
\bea \Gamma&=&|V_{cl}|^2
\frac{G_F^{\,2}}{8\pi^4}\frac{M'^2}{M} \int\sqrt{\omega^2-1}\, {\cal
L}^{\alpha\beta}(q) {\cal H}_{\alpha\beta}(P,P')\,d\omega,
\eea 
where $|V_{cl}|$ is the modulus of the corresponding CKM matrix
element for a semileptonic $c\to l$ decay ($|V_{cs}|=0.97345$ and
$|V_{cd}|=0.2252$~\cite{pdg10}), $G_F= 1.16637(1)\times
10^{-11}$\,MeV$^{-2}$~\cite{pdg10} is the Fermi decay constant, $P,M$
($P',M'$) are the four-momentum and mass of the initial (final)
baryon, $q=P-P'$ and $\omega$ is the product of the initial and final
baryon four-velocities $\omega=v\cdot v'=\frac{P}{M} \cdot
\frac{P'}{M'}=\frac{M^2+M'^2-q^2}{2MM'}$. In the decay, $\omega$
ranges from $\omega=1$, corresponding to zero recoil of the final
baryon, to a maximum value that, neglecting the neutrino mass, is
given by $\omega=\omega_{\rm max}= \frac{M^2 + M'^2-m^2}{2MM'}$, which
depends on the transition and where $m$ is the final charged lepton
mass. Finally ${\cal L}^{\alpha\beta}(q)$ is the leptonic tensor after
integrating in the lepton momenta. It can be cast as
\bea
{\cal L}^{\alpha\beta}(q)=A(q^2)\,g^{\alpha\beta}+
B(q^2)\,\frac{q^\alpha q^\beta}{q^2},
\label{eq:lt}
\eea
where  explicit expressions for the scalar functions $A(q^2)$ and
$B(q^2)$ can be found in Eqs.~(3) and (4) of Ref.~\cite{Albertus:2011xz}.

The hadron tensor ${\cal H}_{\alpha\beta}(P,P')$ is given by
\begin{eqnarray}
{\cal H}^{\alpha\beta}(P,P') &=& \frac{1}{2J+1} \sum_{r,r'}  
 \big\langle B', r'\
\vec{P}^{\,\prime}\big| J_{cl}^\alpha(0)\big| B, r\ \vec{P}   \big\rangle 
\ \big\langle B', r'\ 
\vec{P}^{\,\prime}\big|J_{cl}^\beta(0) \big|  B, r\ \vec{P} \big\rangle^*,
\label{eq:wmunu}
\end{eqnarray}
with $J$ the initial baryon spin, $\big|B, r\ \vec P\big\rangle\,
\left(\big|B', r'\ \vec{P}\,'\big\rangle\right)$ the initial (final)
baryon state with three-momentum $\vec P$ ($\vec{P}\,'$) and spin
third component $r$ ($r'$) in its center of mass
frame. Baryon states are normalized such that 
$  \big\langle B, r'\ \vec{P}'\, |\,B, r \ \vec{P} \big\rangle =
  2E\,(2\pi)^3 \,\delta_{rr'}\, \delta^3 (\vec{P}-\vec{P}^{\,\prime})
 $,
  with $E$ the baryon energy for three-momentum $\vec P$.
 Our
states are constructed in Appendix~A of Ref.~\cite{ahn12}. Finally,
$J_{cl}^\mu(0)=\bar\Psi_{l}(0)\gamma^\mu(1-\gamma_5)\Psi_c(0)$ is the
$c\to l$ charged weak current.

For the actual calculation of the decay width we parameterize the hadronic 
matrix elements  in terms of form factors, which are functions of $\omega$
 or equivalently of $q^2$. The different form factor
decomposition that we use are given in the following.
\begin{enumerate} 
\item $\ 1/2 \to 1/2$ transitions.\\ 
Here we take the commonly used decomposition in terms of  three vector
  $F_1,\,F_2,\,F_3$ and three axial $G_1,\,G_2,\,G_3$ form factors
\begin{eqnarray}
\label{eq:1212}
\big\langle B'(1/2), r'\ \vec{P}^{\,\prime}\left|\,
J_{cl}^\mu(0) \right| B(1/2), r\ \vec{P}
\big\rangle& =& {\bar u}^{B'}_{r'}(\vec{P}^{\,\prime})\Big\{
\gamma^\mu\left[F_1(\omega)-\gamma_5 G_1(\omega)\right]+ v^\mu\left[F_2(\omega)-\gamma_5
G_2(\omega)\right]\nonumber\\
&&\hspace{1.5cm}+v'^\mu\left[F_3(\omega)-\gamma_5 G_3(\omega)
\right]\Big\}u^{B}_r(\vec{P}\,).
\end{eqnarray}
 The $u_{r}$ are Dirac spinors normalized as $({ u}_{r'})^\dagger u_r
 = 2E\,\delta_{r r'}$.  

\item $\ 1/2 \to 3/2$ transitions.\\
In this case we follow Llewellyn
 Smith~\cite{Llewellyn Smith:1971zm} to write
\begin{eqnarray}
\label{eq:1232}
&&\hspace{-1cm}\big\langle B'(3/2),r'\vec P'\,|\,\overline 
\Psi_l(0)\gamma^\mu(1-\gamma_5)\Psi_c(0)\,|\,B(1/2),r\,
\vec P\,\big\rangle=
~\bar{u}^{B'}_{\lambda\,r'}(\vec{P}\,')\,\Gamma^{\lambda\mu}(P,P')\,
u^{B}_r(\vec{P}\,),
\nonumber\\
\Gamma^{\lambda\mu}(P,P')=&&
\left[\frac{C_3^V}{M}(g^{\lambda\,\mu}q
\hspace{-.15cm}/\,
-q^\lambda\gamma^\mu)+\frac{C_4^V}{M^2}(g^{\lambda\,\mu}q\cdot
P'-q^\lambda
P'^\mu)+\frac{C_5^V}{M^2}(g^{\lambda\,\mu}q\cdot P-q^\lambda
P^\mu)+C_6^Vg^{\lambda\,\mu}\right]\gamma_5\nonumber\\
&&+\left[\frac{C_3^A}{M}(g^{\lambda\,\mu}q
\hspace{-.15cm}/\,
-q^\lambda\gamma^\mu)+\frac{C_4^A}{M^2}(g^{\lambda\,\mu}q\cdot P'-q^\lambda
P'^\mu)+{C_5^A}g^{\lambda\,\mu}+\frac{C_6^A}{M^2}
q^\lambda q^\mu\right].
\end{eqnarray}
Here $u^{B'}_{\lambda\,r'}$ is the Rarita-Schwinger spinor of the final spin
3/2 baryon normalized such that $(u_{\lambda\,r'}^{B'})^{\dagger}
u^{B'\,\lambda}_r = -2E'\,\delta_{rr'}$, and we have four vector
($C^V_{3,4,5,6}(\omega)$) and four axial ($C^A_{3,4,5,6}(\omega)$) form
factors. Within our model we shall have that 
$C^V_{5}(\omega)=C^V_{6}(\omega)=C^A_{3}(\omega)=0$.
 
\item $3/2\to 1/2$ transitions.\\
Similar to the case before we use
\begin{eqnarray}
&&\hspace{-1.5cm}\left\langle B'(1/2), r'\ \vec{P}^{\,\prime}\left|\,\overline \Psi_{l'}(0)\gamma^\mu(1-\gamma_5)
\Psi_c(0)
 \right| B(3/2), r\ \vec{P}
\right\rangle =\nonumber\\&&
(\bar{u}^{B}_{\lambda\,r}(\vec{P}\,)\tilde\Gamma^{\lambda\,\mu}(P',P)
u^{B'}_{r'}(\vec{P}\,'))^*
=\bar
u^{B'}_{r'}(\vec{P}\,')\gamma^0(\tilde\Gamma^{\lambda\,\mu}(P',P))^\dagger\gamma^0
{u}^{B}_{\lambda\,r}(\vec{P}),\nonumber\\
%
\tilde\Gamma^{\lambda\,\mu}(P',P)&=&
\big(-\frac{C_3^V}{M'}(g^{\lambda\,\mu}q
\hspace{-.15cm}/\,
-q^\lambda\gamma^\mu)-\frac{C_4^V}{M'^2}(g^{\lambda\,\mu}q\cdot P-q^\lambda
P^\mu)-\frac{C_5^V}{M'^2}(g^{\lambda\,\mu}q\cdot P'-q^\lambda
P'^\mu)+C_6^Vg^{\lambda\,\mu}\big)\gamma_5\nonumber\\&&
+\big(-\frac{C_3^A}{M'}(g^{\lambda\,\mu}q
\hspace{-.15cm}/\,
-q^\lambda\gamma^\mu)-\frac{C_4^A}{M'^2}(g^{\lambda\,\mu}q\cdot P-q^\lambda
P^\mu)+{C_5^A}g^{\lambda\,\mu}+\frac{C_6^A}{M'^2}
q^\lambda q^\mu\big).
\end{eqnarray} 
Again, and within our model, we shall have that 
$C^V_{5}(\omega)=C^V_{6}(\omega)=C^A_{3}(\omega)=0$.

\item $\ 3/2\to3/2$ transitions.\\
A form factor decomposition for  $3/2\to3/2$ can be found in
Ref.~\cite{Faessler:2009xn} where a total of 7 vector plus 7 axial form factors
are needed. In this case we do not evaluate the form factors but work directly
with the vector and axial matrix elements.
\end{enumerate}

Expressions relating form factors to weak current matrix elements can be found in Appendix B of Ref.~\cite{ahn12}.

Heavy Quark Spin Symmetry (HQSS) imposes constraints on the form factors. These
 constraints have been deduced in Ref.~\cite{ahn12}  using the Trace
  Formalism~\cite{Falk:1990yz,MWbook} by requiring invariance under separate
  bottom and charm spin rotations. Though these relations are strictly valid in
  the limit of very large heavy quark mass and near zero recoil of the final
  baryon they turn out to be reasonable accurate for the whole available phase
  space. 

\section{Results}

The results we obtain for the semileptonic decay widths of $cb$
baryons are presented in Tables~\ref{tab:resctos} ($c\to s$ decays)
and \ref{tab:resctod} ($c\to d$ decays).  We show between parentheses
the results obtained ignoring configuration mixing in the spin-1/2
$cb$ initial baryons. 
Due to the finite value of the heavy quark masses, the hyperfine
interaction between the light quark and any of the heavy quarks can
admix both $S$=0 and 1 components into the wave function for total
spin-1/2 states. Thus, the actual physical spin-1/2 $cb$ baryons 
that we call
$\Xi_{cb}^{(1)},\,\Xi_{cb}^{(2)}$ and
$\Omega_{cb}^{(1)},\,\Omega_{cb}^{(2)}$, and that were obtained in  
Ref.~\cite{Albertus:2009ww}, are
admixtures of the $\Xi_{cb},\,\Xi'_{cb}$
($\Omega_{cb},\,\Omega'_{cb}$) states where the $c$ and $b$ quarks are
coupled to well defined total spin $S$=1,0.  
While masses are not very
sensitive to hyperfine mixing, it was pointed out
by Roberts and Pervin~\cite{pervin1} that hyperfine mixing could
greatly affect the decay widths of doubly heavy $cb$ baryons.  This
assertion was checked in Ref.~\cite{pervin2} where Roberts and Pervin
found that hyperfine mixing in the $cb$ states has a tremendous impact
on doubly heavy baryon $b\to c$ semileptonic decay widths. These
results were qualitatively confirmed by our own calculation in
Ref.~\cite{Albertus:2009ww}. We further investigated the role of
hyperfine mixing in electromagnetic transitions~\cite{Albertus:2010hi}
finding again large corrections to the decay widths. A similar study
was conducted by Branz et al. in Ref.~\cite{Branz:2010pq}.  We expected
configuration mixing should also play an important role for $c\to s,d$
semileptonic decay of $cb$ baryons.  Indeed, we find  that  configuration
 mixing has an
important effect when the two light quarks in the final state couple
to total spin 0.

\begin{table}[h!!!]
\begin{tabular}{llccc}
&\multicolumn{4}{c}{$\Gamma \ [10^{-14}\,{\rm GeV}]$}\\
&{This work}&\cite{sanchis95}&\cite{Faessler:2001mr}&\cite{Kiselev:2001fw}\\
\hline
$\Xi^{(1)\,+}_{cbu}\to\Xi^0_b\, e^+\nu_e$& 3.74 (3.45)&(3.4)\\
$\Xi^{(2)\,+}_{cbu}\to\Xi^0_b\, e^+\nu_e$& 2.65 (2.87)\\
$\Xi^{(1)\,+}_{cbu}\to\Xi'^0_b\, e^+\nu_e$& 3.88
(1.66)&&$2.44\div3.28^\dagger$\\
$\Xi^{(2)\,+}_{cbu}\to\Xi'^0_b\, e^+\nu_e$&1.95 (3.91)\\
$\Xi^{(1)\,+}_{cbu}\to\Xi^{*\,0}_b\, e^+\nu_e$& 1.52 (3.45)\\
$\Xi^{(2)\,+}_{cbu}\to\Xi^{*\,0}_b\, e^+\nu_e$& 2.67 (1.02)\\
$\Xi^{(2)\,+}_{cbu}\to\Xi^0_b\, e^+\nu_e+\Xi'^0_b\, e^+\nu_e+\Xi^{*\,0}_b\, e^+\nu_e$& 7.27 (7.80)
&&&$(9.7\pm1.3)^*$\\
$\Xi^{*\,+}_{cbu}\to\Xi^{0}_b\, e^+\nu_e$&  4.08\\
$\Xi^{*\,+}_{cbu}\to\Xi'^{0}_b\, e^+\nu_e$&0.747\\
$\Xi^{*\,+}_{cbu}\to\Xi^{*\,0}_b\, e^+\nu_e$& 5.03\\\hline
\end{tabular}\hspace{.5cm}
\begin{tabular}{ll}
&{\hspace*{.375cm}$\Gamma \ [10^{-14}\,{\rm GeV}]$}\\
\hline
$\Omega^{(1)\,0}_{cbs}\to\Omega^-_b\, e^+\nu_e$&\hspace*{.5cm} 7.21 (3.12)\\
$\Omega^{(2)\,0}_{cbs}\to\Omega^-_b\, e^+\nu_e$&\hspace*{.5cm} 3.49 (7.12)\\
$\Omega^{(1)\,0}_{cbs}\to\Omega^{*\,-}_b\, e^+\nu_e$&\hspace*{.5cm} 2.98 (6.90)\\
$\Omega^{(2)\,0}_{cbs}\to\Omega^{*\,-}_b\, e^+\nu_e$&\hspace*{.5cm} 5.50 (2.07)\\
$\Omega^{*\,0}_{cbs}\to\Omega^-_b\, e^+\nu_e$&\hspace*{.5cm} 1.35\\
$\Omega^{*\,0}_{cbs}\to\Omega^{*\,-}_b\, e^+\nu_e$&\hspace*{.5cm} 10.2\\\hline
\end{tabular}
\caption{$\Gamma$ decay widths for  $c\to s$ decays. Results where configuration
mixing is not considered are shown in between parentheses. In this latter case
 the
$\Xi^{(1)}_{cb},\ \Xi^{(2)}_{cb}$ baryons in the table should be interpreted
respectively as the $\Xi'_{cb},\ \Xi_{cb}$ states. The result with a
$\dagger$ corresponds to the decay of the $\widehat \Xi_{cb}$
state (see main text). The result 
with an ${\ast}$ is our
estimate from the total decay width and the branching ratio
given in~\cite{Kiselev:2001fw}. Similar results are obtained for decays into
$\mu^+\nu_\mu$.} 
\label{tab:resctos}
\end{table}
\begin{table}[h!!!]
\begin{tabular}{ll}
&{\hspace*{.5cm}$\Gamma \ [10^{-14}\,{\rm GeV}]$}\\
\hline
$\Xi^{(1)\,+}_{cbu}\to\Lambda^0_b\, e^+\nu_e$&\hspace*{.5cm} 0.219 (0.196)\\
$\Xi^{(2)\,+}_{cbu}\to\Lambda^0_b\, e^+\nu_e$&\hspace*{.5cm}  0.136 (0.154)\\
$\Xi^{(1)\,+}_{cbu}\to\Sigma^0_b\, e^+\nu_e$&\hspace*{.5cm}  0.198 (0.0814)\\
$\Xi^{(2)\,+}_{cbu}\to\Sigma^0_b\, e^+\nu_e$ &\hspace*{.5cm} 0.110 (0.217)\\
$\Xi^{(1)\,+}_{cbu}\to\Sigma^{*\,0}_b\, e^+\nu_e$&\hspace*{.5cm}  0.0807 (0.184)\\
$\Xi^{(2)\,+}_{cbu}\to\Sigma^{*\,0}_b\, e^+\nu_e$&\hspace*{.5cm}  0.147 (0.0556)\\
$\Xi^{*\,+}_{cbu}\to\Lambda^{0}_b\, e^+\nu_e$&\hspace*{.5cm}   0.235\\
$\Xi^{*\,+}_{cbu}\to\Sigma^{0}_b\, e^+\nu_e$&\hspace*{.5cm}  0.0399\\
$\Xi^{*\,+}_{cbu}\to\Sigma^{*\,0}_b\, e^+\nu_e$&\hspace*{.5cm}  0.246\\\hline
\end{tabular}\hspace{2cm}
\begin{tabular}{ll}
&{\hspace*{.5cm}$\Gamma \ [10^{-14}\,{\rm GeV}]$}\\
\hline
$\Omega^{(1)\,0}_{cbs}\to\Xi^-_b\, e^+\nu_e$&\hspace*{.5cm}  0.179 (0.164)\\
$\Omega^{(2)\,0}_{cbs}\to\Xi^-_b\, e^+\nu_e$&\hspace*{.5cm}  0.120 (0.133)\\
$\Omega^{(1)\,0}_{cbs}\to\Xi'^-_b\, e^+\nu_e$&\hspace*{.5cm}  0.169 (0.0702)\\
$\Omega^{(2)\,0}_{cbs}\to\Xi'^-_b\, e^+\nu_e$&\hspace*{.5cm}  0.0908 (0.182)\\
$\Omega^{(1)\,0}_{cbs}\to\Xi^{*\,-}_b\, e^+\nu_e$&\hspace*{.5cm}  0.0690 (0.160)\\
$\Omega^{(2)\,0}_{cbs}\to\Xi^{*\,-}_b\, e^+\nu_e$&\hspace*{.5cm}  0.130 (0.0487)\\
$\Omega^{*\,0}_{cbs}\to\Xi^-_b\, e^+\nu_e$&\hspace*{.5cm}  0.196\\
$\Omega^{*\,0}_{cbs}\to\Xi'^-_b\, e^+\nu_e$&\hspace*{.5cm}  0.0336\\
$\Omega^{*\,0}_{cbs}\to\Xi^{*\,-}_b\, e^+\nu_e$&\hspace*{.5cm}  0.223\\\hline
\end{tabular}\caption{$\Gamma$ decay widths for  
$c\to d$ decays. In between parentheses
we show the results without configuration mixing. Similar results are obtained
for decays into
$\mu^+\nu_\mu$.}
\label{tab:resctod}
\end{table}

\begin{figure}
\begin{center}
\rotatebox{270}{\resizebox{!}{10cm}{\includegraphics{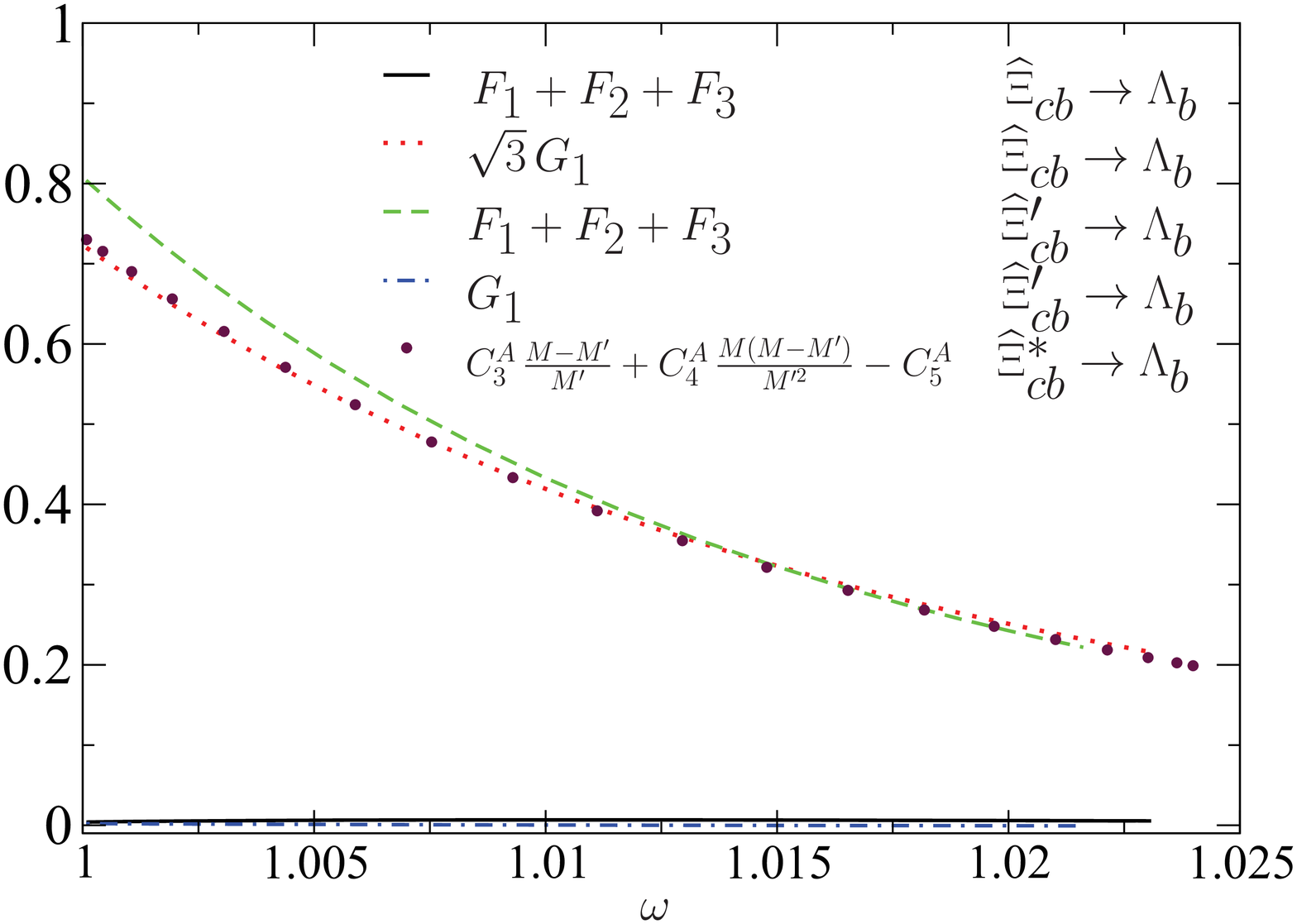}}}
\caption{Test of HQSS constraints: Different combinations of form
  factors obtained in this work for several transitions with a
  $\Lambda_b$ in the final state ($S'_{\rm light}=0$).  For the
  calculation we have taken the masses of the
  $\widehat\Xi_{cb},\widehat\Xi'_{cb}$ to be the masses of the
  physical states $\Xi^{(1)}_{cb},\Xi^{(2)}_{cb}$.  For very large heavy
  quark masses, HQSS predicts that the combination of form factors in 
  the second, third and fourth lines
  should be equal, while for the first and fifth line they should be zero.
 For other transitions see Ref.~\cite{ahn12}.\vspace*{.5cm}}
\label{fig:xitola}
\end{center}
\end{figure}

In Fig.~\ref{fig:xitola} we check that our calculation respects the
constraints on the form factors deduced in Ref.~\cite{ahn12} using HQSS. Those constraints have been 
deduced for the  $\widehat B_{cb}=-\frac{\sqrt3}2B'_{cb}+\frac12B_{cb}$ and
$\widehat B'_{cb}=\frac12B'_{cb}+\frac{\sqrt3}2B_{cb}$, where the spins of the
$c$ and light quark couple to total spin 1 and 0 respectively. These hatted
states are very close to our physical  $B^{(1)}_{cb}$ and $B^{(2)}_{cb}$ states.
 One
sees deviations at the 10\% level near zero recoil. Those deviation stem from
 corrections in the inverse of the heavy quark
masses. In fact the constraints
are satisfied to that level of accuracy over the whole $\omega$ range
accessible in the decays. We found similar deviations in our recent
study of the $c\to s,d$ decays of double charmed baryons in
Ref.~\cite{Albertus:2011xz}, where we explicitly showed these
discrepancies tend to disappear when the mass of the heavy quark is
made arbitrarily large.  

Besides, in Ref.~\cite{ahn12}, with the use of the HQSS relations and 
assuming $M_{B_{cb}}=
M_{B'_{cb}} = M_{B^*_{cb}}$ and $M_{B_{b}}= M_{B'_{b}}= M_{B^*_{b}}$,
we have made model independent, though approximate, predictions for
ratios of $c\to s,d$ decay widths of $cb$ doubly heavy baryons. Our values 
for those ratios agree with the
HQSS motivated predictions at the level of 10\% in most of the
cases. We expect those predictions to hold to that level of accuracy
in other approaches.
\begin{acknowledgments}
  This research was supported by DGI and FEDER funds, under contracts
   FPA2010-21750-C02-02, FIS2011-28853-C02-02, and the Spanish
  Consolider-Ingenio 2010 Programme CPAN (CSD2007-00042),  by Generalitat
  Valenciana under contract PROMETEO/20090090 and by the EU
  HadronPhysics2 project, grant agreement no. 227431. C. A. thanks a Juan de 
  la Cierva contract from the
Spanish  Ministerio de Educaci\'on y Ciencia.
\end{acknowledgments}

\end{document}